\documentclass[prl,twocolumn,showpacs,superscriptaddress]{revtex4}

\usepackage{ulem}
\usepackage{amsmath}
\usepackage{graphicx,color}
\usepackage{dcolumn}
\usepackage{bm}
\begin{document}
\title{Manipulation of Non-classical Atomic Spin States}

\author{Tetsushi Takano}
\affiliation{Department of Physics, Graduate School of Science, Kyoto University, Kyoto 606-8502, Japan}
\author{Shin-Ichi-Ro Tanaka}
\affiliation{Department of Physics, Graduate School of Science, Kyoto University, Kyoto 606-8502, Japan}
\author{Ryo Namiki}
\affiliation{Department of Physics, Graduate School of Science, Kyoto University, Kyoto 606-8502, Japan}
\author{Yoshiro Takahashi}
\affiliation{Department of Physics, Graduate School of Science, Kyoto University, Kyoto 606-8502, Japan}
\affiliation{CREST, JST, 4-1-8 Honcho Kawaguchi, Saitama 332-0012, Japan}

\date{\today}

\begin{abstract}
We report successful manipulation of non-classical atomic spin states.
We generate squeezed spin states by a spin
 quantum nondemolition measurement, and apply an off-resonant circularly-polarized light pulse to the atoms.
By changing the pulse duration, we have clearly observed a rotation of anisotropic quantum noise distribution in good contrast with the case of manipulation of 
a coherent spin state where the quantum noise distribution is always isotropic.  
This is an important step for quantum state tomography, quantum swapping, and precision spectroscopic measurement.

\end{abstract}
\maketitle
\parskip=0pt

A quantum noise is a central subject in quantum physics.
While a quantum noise of a light field has been well studied in quantum optics \cite{NAT390-579},
there has been a growing interest in a quantum noise of an atomic spin ensemble \cite{PRA47-5138}, because the squeezed quantum noise can improve the measurement precision of an atomic spectroscopy such as an atomic clock \cite{PRA47-3554,PRA75-033803,PRL96-043001} 
and is also recognized as an important resource for the continuous variable quantum information processing \cite{PRL84-4232,PRL88-243602,PRA62-033809,PRA70-044304,PRA74-011802R}.
Since the theoretical proposals for generating the so-called squeezed spin state \cite{PRA47-5138} based on a spin quantum nondemolition measurement (spin-QNDM) using the Faraday rotation (FR) interaction \cite{PRA60-4974,EPL42-481}, experimental challenge to reduce quantum spin noise has been actively studied in the past decade \cite{PRL85-1594,NAT413-400}.   
Quite recently the important progresses have been reported 
on the realization of the squeezed spin state via a spin-QNDM for the nuclear spin one-half of cold ytterbium ($^{171}\mathrm{Yb}$) atoms \cite{PRL102-033601} and the hyperfine-clock transitions of cold alkali-metal atoms \cite{arx0810-2582,arx0810-3545}.
See Fig. 1.
While the rotation of a classical spin state, the so-called coherent spin state, in the Bloch sphere is a basic operation widely used in many applications \cite{NAT432-482}, 
no results on the manipulation of a non-classical spin state have been reported until now.
Therefore, the next crucial step especially for quantum information processing \cite{PRA78-010307R,arx0905-1197} and precision measurement is to manipulate the squeezed spin state.
So far, a quantum memory was realized by applying a magnetic feedback to coherent spin states \cite{NAT432-482}, and 
a deterministic spin squeezing by a quantum feedback was proposed \cite{PRA69-032109}.
Since the important feature characterizing the squeezed spin state is the anisotropy of the quantum spin noise and not the average spin value, 
essentially important is the manipulation of the anisotropic quantum spin noise distribution of the squeezed spin state without decoherence \cite{Vuletic}. 
This is in good contrast with the manipulation of a coherent spin state where the quantum noise distribution is always isotropic \cite{NAT432-482}. 
See Fig. 1 (c) C-F.
\begin{figure}
\includegraphics[width=8.5cm]{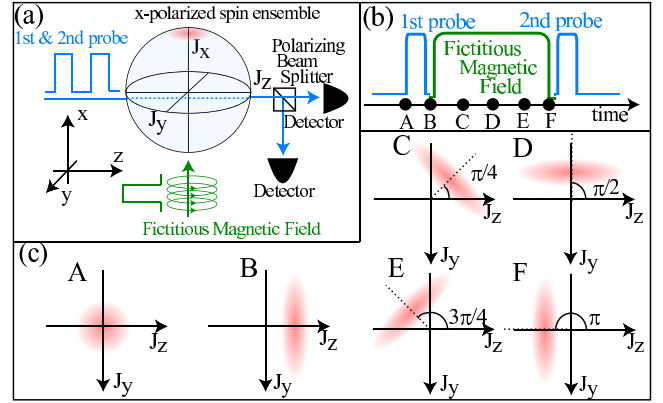}
\caption{(color online) (a) Experimental setup.
(b) Experimental time sequence.
Two successive linearly-probe pulses pass through the $x$-polarized $^{171}\mathrm{Yb}$ atoms 
and
the polarizations are measured.
The spin is rotated about the $x$-axis by a fictitious magnetic field applied on the atoms in the slot of the two-pulse train.
(c) A schematic view of the behaviors of the spin distributions. A: coherent spin state, B: squeezed spin state, and C-F: rotated squeezed spin states. 
Note that the squeezed spin state is prepared by conditioning on the first measurement outcome, and the position in the $J_z$ direction on B varies depending on the outcome.
 }
\end{figure}

In this Letter, we report a successful demonstration of a rotational manipulation of the squeezed spin state of the nuclear spin one-half of cold $^{171}\mathrm{Yb}$ atoms via a fictitious magnetic field produced by an off-resonant circularly-polarized light \cite{PRA5-968}. A schematic view of the manipulated spin state is shown in Fig. 1. 
By changing the pulse duration of the fictitious magnetic field, we have clearly observed a rotation of anisotropic quantum noise distribution of the squeezed spin states up to the angle of $\pi$ by measuring the quantum correlations between the two-successively applied light polarizations, the first light pulse for the spin-QNDM and the second for the verification applied after the rotation. 
In particular, the squeezed spin state associated with the population difference for the $z-$spin component (Fig. 1(c) B) can be successfully converted into a phase-squeezed spin state (Fig. 1(c) D) .
It is also confirmed that the initial squeezing level is retrieved after a $\pi$ rotation
owing to the quick rotation with the angular velocity of typically 0.4 rad/$\mu$s and a long coherence time of the  nuclear spin \cite{PRL102-033601}.

Let us define the collective spin operator of the atoms 
$\vec{{{J}}} =({J_x},{J_y},{J_z})=\sum_{i=1}^{N_A} \vec j_i$, where $\vec{j}_i$ is a spin operator of a single atom and $N_A$ is the number of the atoms \cite{PRA47-5138}. 
The Stokes operator of a pulsed light $\vec {{S}} =( {S_x}, {S_y}, {S_z})$
is defined by ${S_x} = 1/2\int _0^t(a_+^{\dagger}a_-+a_-^{\dagger}a_+)dT$, 
${S_y} = 1/(2i)\int _0^t(a_+^{\dagger}a_--a_-^{\dagger}a_+)dT$, and ${S_z} = 1/2\int _0^t(a_+^{\dagger}a_+-a_-^{\dagger}a_-)dT$, 
where 
$t$ is the pulse duration, and $a_{\pm}$ is the
annihilation operator of $\sigma _{\pm}$ circular polarization mode, respectively \cite{PRL85-5643}.
The Hamiltonian of the FR interaction is given by
$
H_{int}=\alpha {S_z}{J_z},
$
where $\alpha$ is a real constant and $z$ means the propagation direction of the light \cite{PRA60-4974,EPL42-481}.
To simplify expressions, let us introduce the normalized operator 
$\tilde {\vec J}\equiv \vec J / \sqrt{J}$ and $\tilde {\vec S}\equiv \vec S / \sqrt{S}$, where $J=N_A/2$ and $S=N_L/2$
 and $N_L$ is the mean photon number of the pulse. 
In the case of spin-QNDM,
the initial states of the light and atoms
are prepared in the $x$-polarized coherent states, 
namely,
 $
\tilde J_x \simeq \sqrt{J},
$
$
\langle \tilde J_y \rangle = \langle \tilde J_z \rangle=0,
$
$
\tilde S_x \simeq \sqrt{S},
$
and
$
\langle \tilde S_y \rangle = \langle \tilde S_z \rangle=0.
$
The variances of these operators are
$
V(\tilde J_y )= V(\tilde J_z ) = V(\tilde S_y )= V(\tilde S_z ) = 1/2,
$
where $ V(X) \equiv \langle X ^2\rangle -  \langle X \rangle ^2$ denotes the variance of an operator $X$.
The FR interaction transforms the operators as 
$
\tilde S_y \to\tilde S_y  + \kappa\tilde  J_z$, $\tilde S_z \to\tilde S_z $,
$\tilde J_y \to\tilde J_y  + \kappa \tilde S_z $, and $\tilde J_z \to\tilde J_z ,
$ where the interaction strength $\kappa $ is given by $\kappa \equiv \alpha t\sqrt{JS}$.
Note that this interaction satisfies a back-action evading condition $[H_{int}, \tilde{J_z}] = 0$, and makes a quantum correlation between $\tilde J_z$ and $\tilde S_y$. Thereby, the measurement of $\tilde S_y $ will essentially project the spin state into an eigenstate of $\tilde J_z$ and the variance of $\tilde J_z$ is squeezed \cite{PRA60-4974,EPL42-481,PRL102-033601}.

The fictitious magnetic field effect  \cite{PRA5-968} is easily understood from the FR interaction given by
$
H_{int}=\alpha' {S_x}{J_x},
$
where $\alpha'$ is again a real constant and now the the light for the fictitious magnetic field 
propagates in the $x$-direction.
The application of a strong circularly-polarized light along the $x$-axis ($S_x \simeq S \equiv N_L/2$)
results in the rotation of the spin ensemble about the $x$-axis, such that,
 $
J_y (t) = J_y (0) \cos (\alpha t S) - J_z (0)\sin (\alpha t S)
$
and
$
J_z (t) = J_z (0) \cos (\alpha t S) + J_y (0)\sin (\alpha t S)
$.

In our experiment, a linearly-polarized probe pulse which is represented by $(\tilde{S}_{1,y}^{(i)},\tilde{S}_{1,z}^{(i)})$ 
interacts with an spin ensemble $(\tilde{J}_{y}^{(i)},\tilde{J}_{z}^{(i)})$, at first.
See Fig. 1 (b) for the time sequence.
This interaction transforms the operators as
$
\tilde S_{1,y} ^{(1)}=\tilde S_{1,y} ^{(i)}+ \kappa\tilde  J_z ^{(i)}$, $\tilde S_{1,z} ^{(1)}=\tilde S_{1,z}  ^{(i)}$,
$\tilde J_y ^{(1)}=\tilde J_y ^{(i)} + \kappa \tilde S_{1,z} ^{(i)}$, and $\tilde J_z ^{(1)}=\tilde J_z^{(i)}
$.
Secondly, the spin ensemble is rotated by $\phi$ about the $x$ axis by a fictitious magnetic field
 generated by a circularly-polarized light pulse which propagates in the $x$ direction, so that,
$\tilde J_z ^{(2)}= \tilde J_z ^{(1)} \cos \phi  +  \tilde J_y ^{(1)} \sin \phi  $.
Then, another linearly-polarized pulse represented by $(\tilde{S}_{2,y}^{(i)},\tilde{S}_{2,z}^{(i)})$ goes through the spin ensemble, yielding 
$
\tilde S_{2,y} ^{(3)}=\tilde S_{2,y} ^{(i)}+ \kappa\tilde  J_z ^{(2)}$ and $\tilde S_{2,z} ^{(3)}=\tilde S_{2,z} ^{(i)}$.
Finally, we measure  $\tilde S_{1,y} ^{(1)}$ and $\tilde S_{2,y} ^{(3)}$ as well as the quantum correlation between them.
If all of the initial states are coherent states and the loss is negligible, we have 
\begin{align}
V_1 \equiv V(\tilde S_{1,y} ^{(1)}) &= (1 + \kappa^2 )/2 \\
V_2 \equiv V(\tilde S_{2,y} ^{(3)}) &= (1 + \kappa^2 + \kappa^4 \sin ^2 \phi)/2 \\
V_\pm \equiv V(\frac{\tilde S_{1,y} ^{(1)}\pm \tilde S_{2,y} ^{(3)}}{\sqrt{2}})&=\frac{2+\kappa ^2(2\pm 2\cos \phi )+\kappa ^4\sin ^2 \phi }{4}.
\end{align}
Note that the case of $\phi = 0$ corresponds to the previous spin-QNDM \cite{PRL102-033601}.
Equation (3) implies that 
the correlation can be controlled by the spin rotation.
For example, if we change the rotation angle $\phi$ like $0 \to \pi /2 \to \pi$,
$V_+$ becomes $(1+ 2\kappa ^2)/2 \to (2+2\kappa ^2 + \kappa^4)/4  \to 1/2$,
whereas $V_-$ becomes $1/2 \to (2+2\kappa ^2 + \kappa^4)/4  \to (1+ 2\kappa ^2)/2 $.
To experimentally demonstrate these behaviors by changing the rotation angle $\phi$ is, therefore, one of the main results of this work, which is later shown in Fig. 2.  
\par
The experiment is done with cold $^{171}\mathrm{Yb}$ atoms released from a magneto-optical trap (MOT) \cite{PRL102-033601}. 
See Fig. 1(a) and (b) for the schematic setup and sequence of the experiment. 
At first, typically $10^6$ atoms are loaded in the MOT in 1 $\mathrm{s}$ and are released in the next 250 $\mathrm{\mu s}$.
Secondly, the atoms are polarized in the $x$ direction by the circularly-polarized resonant pumping pulse whose width is 9 $\mu s$.
Then, two linearly-polarized probe pulses focused on the atomic region pass through the atoms in the $z$ direction and go into the polarization detector. After the detection of the second pulse, the atoms are recaptured by the MOT.
The pulses have the same width of 100 $\mathrm{ns}$ and the interval between them is 10 $\mathrm{\mu s}$. 
The FR angle $\alpha t J /2$ is about $200$ $\mathrm{mrad}$ which corresponds to $J\simeq 5.0\times 10^5$.
The wavelength of the probe beam is about 399 nm and the frequency is locked at the center of the two hyperfine splitting 
of the $^1\mathrm{S}_0\leftrightarrow{^1\mathrm{P}_1}$ transitions of $^{171}$Yb. 
 $\kappa$ is calculated as
\begin{equation}
\kappa = \frac{\Gamma \sigma_0 \sqrt{SJ}}{3\pi  w_0 ^2}\big( \frac{\delta }{\delta ^2+ (\Gamma /2)^2}-\frac{\delta +\delta _0}{(\delta +\delta _0) ^2+ (\Gamma /2)^2} \big),
\end{equation}
where $ \Gamma = 2\pi \times  29$ $\mathrm{MHz}$ is the natural full linewidth, $\sigma_0 = 7.6\times 10^{-14}$ $\mathrm{m^2}$ is the photon-absorption cross section of $^{171}\mathrm{Yb}$ atom, $w _0=61$ $\mathrm{\mu m}$ is the beam waist, $\delta =- 2\pi \times 160$ $\mathrm{MHz} $ is the 
detuning from the $^1\mathrm{S}_0\leftrightarrow {^1\mathrm{P}_1}$ ($F'= 1/2$) states, and
$ \delta _0 = 2\pi \times 320$ $\mathrm{MHz}$ is the frequency difference between the $F'=1/2$ and $F'=3/2$ states in the $^1\mathrm{P}_1$ state \cite{APB83-107}.
In our experimental conditions, $\kappa$ is estimated to be 0.63 with $N_L=2.6 \times 10^{6}$, and
 the atomic loss parameter $\epsilon  _A\equiv rt/2$ is $6.7 \times 10^{-2}$, where $r$ is the absorption rate \cite{PRL85-5643}. 
\par
Between the two interactions due to the probe pulses, the fictitious magnetic field is applied by the circularly-polarized light pulse propagating in the $x$ direction. The amount of the rotation angle $\phi$ is adjusted by the width of this fictitious magnetic field pulse.
This light pulse is generated from a frequency-doubled fiber-laser \cite{APB92-33} and  the wavelength is adjusted nearly-resonant to the 
$^1\mathrm{S}_0\leftrightarrow {^3\mathrm{P}_1}$ ($F'= 1/2$) transition, at the wavelength of 556 nm.
This transition has the natural linewidth $ \Gamma _{FM}= 2\pi \times  182$ $\mathrm{kHz}$ and
the hyperfine splitting $ \delta _{0(FM)}\simeq 2\pi \times 5.9$ $\mathrm{GHz}$,
and so the condition $\Gamma _{FM}\ll \delta _{FM}\ll \delta _{0(FM)}$ is satisfied in a relatively wide range of the detuning $\delta_{FM}$.
In this condition, we can obtain a large rotational angle with a relatively small loss.
In our experiment, we chose $ \delta _{FM}\simeq 2\pi \times 20$ $\mathrm{MHz}$ which results in large rotation of up to $3\pi$
and the speed of the rotation of about 0.4 rad/$\mu$s for about 30 mW input power. 
The beam cross section of about 6 cm $\times$ 2 cm is much wider than the atomic distribution and the probe beam waist to avoid the decoherence due to the inhomogeneity.
As a result, we have successfully observed about 95$\%$ reversal of the FR signal with a 9-$\mu$s fictitious magnetic field pulse. 
\par
For various rotation angles $\phi$, we have measured about 1300 pairs of the Stokes operators $(\tilde S_{1,y} ^{(1)}, S_{2,y} ^{(3)})$.  
For a single loading of atoms in the MOT, we measure 10 pairs of $(\tilde S_{1,y} ^{(1)}, S_{2,y} ^{(3)})$ by repeating the release-measurement-recapture cycle 10 times. Note that we have compensated a zero bias of the measurement data depending on the orders among the 10 cycles and loading sequences.
\begin{figure}
\includegraphics[width=8.5cm]{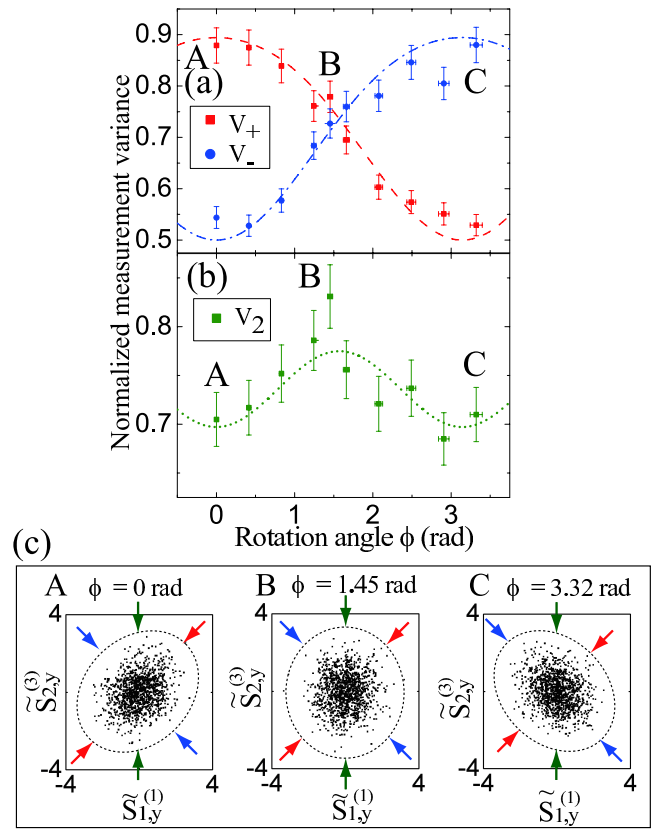}
\caption{(color online)(a) Variance of the sum and difference of the polarizations of the two probe pulses as a function of rotation angle $\phi$,
which are represented by $V_+$ and $V _{-}$, respectively.
As expected from Eq. (3), we have successfully observed an interchange of $V_+$ and $V _{-}$ at rotation angle larger than $\pi /2$.
The horizontal error bar is the fitting error bar.
The vertical error bar is calculated from the statistical error of the variance measurement 
which is written as $\Delta V \equiv \sqrt{\langle V ^2\rangle -  \langle V \rangle ^2}\simeq\sqrt{2/N_m}V$, where $N_m \simeq 1300$ is the number of the measurements.
(b) Variance of the polarization of the second verifying pulse as a function of rotation angle $\phi$.
The variance increases around $\pi /2$, as expected from Eq. (2).
(c) Joint distributions of $\tilde S_{1,y} ^{(1)}$ and $\tilde S_{2,y} ^{(3)}$ for typical rotation angles.
}
\end{figure}

Figure 2 (a) and (b) show the values of $V_\pm$ and  $V_2$, respectively, 
as a function of the rotation angle, while Fig. 2 (c) shows the joint distributions of $\tilde S_{1,y} ^{(1)}$ and $\tilde S_{2,y} ^{(3)}$
for typical rotation angles.
The points with the maximal rotation of 3.3 rad is obtained with a 8-$\mu$s fictitious magnetic field pulse.
As expected from Eq. (3), we have successfully observed an interchange of $V_+$ and $V _{-}$ at rotation angle larger than $\pi /2$.
One can also see that the variance $V_2$ shown in Fig. 2 (b) increases around $ \phi= \pi /2$, as expected from Eq. (2).
This increase results from the back action of the first probe light.
Here, the red dashed curve and the blue dash-dotted curve in Fig. 2 (a) and the green dotted curve in Fig. 2 (b) are theoretical values of the $V_+$ and $V _{-}$ given by Eq. (3) and $V_2$ given by Eq. (2), respectively.
As one can see, the measured variances are almost consistent with the calculated values, which means the successful manipulation of the quantum spin noise with no discernable decoherence effect.
\par
Finally, we show in Fig. 3 the results of the analysis of the data for a squeezed spin state. The measurement of the first pulse induces the squeezing of the atomic spins and the measured value determines the center of the squeezed component $J_z$ (see B of Fig. 1 (c)) \cite{PRA60-4974,EPL42-481,PRL102-033601}.  
By subtracting the shot-by-shot randomness of this center of $J_z$ using the correlated first measurement result, the degree of the squeezing is estimated in the term of the conditional variance \cite{QKD} as,
\begin{align}
V_{cond} &\equiv \min _g V(\tilde S_{2,y} ^{(3)}-g \tilde S_{1,y} ^{(1)}\cos \phi )\notag\\
&= (1 + \kappa^2 + \kappa^4 \sin ^2 \phi - \frac{\kappa ^4}{1+\kappa^2}\cos ^2 \phi)/2,
\end{align}
where the minimum is obtained at $g= \kappa ^2/(1+\kappa^2) $. 
Note that $V_{cond}-1/2$ represents the variance of the squeezed spin component in the case of $\phi = 0$ and $\pi$,
whereas it represents the variance of the anti-squeezed spin component in the case of $\phi = \pi/2$ and $3\pi/2$.
Figure 3 (a) shows $V_{cond}$ as a function of the rotation angle $\phi$. Here, $g = 0.24$ was selected to minimize the sum of the variances all over the measured angles,
$\min _g \sum_\phi  V(\tilde S_{2,y} ^{(3)}-g \tilde S_{1,y} ^{(1)}\cos \phi )$.
The squeezing level at $\phi=0$ is $0.7_{-0.6}^{+0.8}$ dB, and 
upon the rotation up to $\pi/2$, the variance increases.
This growth of the variance originates from the anti-squeezed spin component, and not from the decoherence effect.
This is clear from the observation that the variance around $\phi=\pi$ again becomes smaller than the noise level associated with the coherent spin noise, which also gives $0.7_{-0.6}^{+0.8}$ dB spin squeezing.
 These behaviors correspond to the schematic views of the squeezed spin states shown in Fig. 1 (c). 
The good agreement between the experimental data represented by solid squares and theoretical value of $V_{cond}$ calculated from Eq. (5) represented by the dash-dotted curve is one of the main results of this work,   
from which we claim the success of manipulating the squeezed spin state. 
For reference, we also show in Fig. 3 (b) the variance for the coherent spin state $V_{coh}$.
In this measurement, the first probe pulse for the spin-QNDM is not applied.
The variance of the coherent spin state can be estimated as
\begin{equation}
V_{coh} \equiv V(\tilde S_{2,y} ^{(3)}) =(1 + \kappa^2 )/2.
\end{equation}
As expected, the measured variances are almost independent of $\phi$.
Note that the quantum uncertainty of the coherent spin state is estimated from these values.
\begin{figure}
\includegraphics[width=5.5cm]{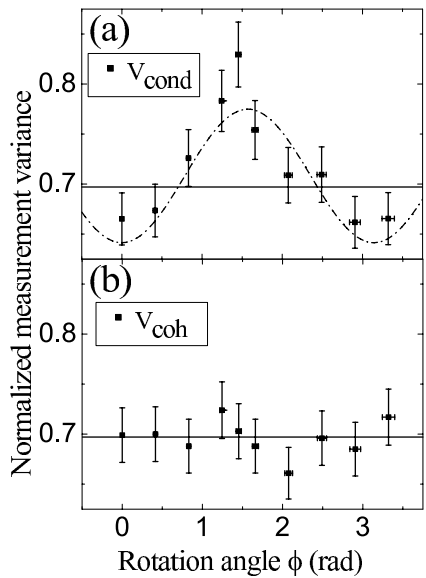}
\caption{Measured variance of the squeezed spin state and coherent spin state as a function of rotation angle $\phi$.
(a) Conditional variances with $g = 0.24$  showing the phase-sensitive quantum-noise behavior of squeezed spin state.
The squeezing of about $0.7_{-0.6}^{+0.8}$ dB below the shot noise level (Solid line) with no rotation of $\phi=0$ is also retrieved after the nearly $\phi= \pi$ rotation.
(b) Variances for the coherent spin state. The measurements are done without the first probe pulse. As expected from Eq. (6), the variance is isotropic. The meaning of the error bars is the same as those in Fig. 2.
}
\end{figure}
\par
In conclusion, we have successfully rotated the phase of the spin ensemble of the nuclear spin one-half of cold $^{171}\mathrm{Yb}$ atoms, and have observed the phase-sensitive quantum noise behavior of a squeezed spin state in the good agreement between the theory and experiment. The rotation by the use of a light field has been executed within a duration of 10 $\mu$s much shorter than the coherence time of the system without inducing discernable effect of decoherence. 
This method will lead us to the dramatical improvement of the performance of the quantum feedback \cite{PRA69-032109} and the realization of the multiple interaction protocols, such as a quantum swapping \cite{PRA78-010307R} or a quantum tomography \cite{arx0905-1197} as well as the precision measurements.

This work was supported by the Grant-in-Aid for Scientific Research of JSPS (No. 18204035) and GCOE Program ``The Next Generation of Physics, Spun from Universality and Emergence'' from MEXT of Japan.
T. T. and R. N. are supported by JSPS Research Fellowships.

\end{document}